\documentclass[12pt]{article}

\usepackage{epsfig,amsfonts,amssymb,newlfont,rotate,float,array}

\DeclareMathAlphabet{\mathitbf}{T1}{cmr}{bx}{it}

\newcommand{\e}{\mathrm e}

\newcommand{\bbox}[1]{\mbox{\boldmath $#1$}} 
\newcommand{\sbbox}[1]{\mbox{\scriptsize\boldmath $#1$}}

\begin{document}
\title{Finite Size Scaling and ``perfect'' actions:\\
the three dimensional Ising model.}
\author{H.~G.~Ballesteros\footnote{\tt hector@lattice.fis.ucm.es}~,~
        L.~A.~Fern\'andez\footnote{\tt laf@lattice.fis.ucm.es}~,\\
        V.~Mart\'{\i}n-Mayor\footnote{\tt victor@lattice.fis.ucm.es}~,~
        A.~Mu\~noz Sudupe\footnote{\tt sudupe@lattice.fis.ucm.es}~,\\
\normalsize \it Departamento de F\'{\i}sica Te\'orica I, 
        Facultad de CC. F\'{\i}sicas,\\
\normalsize \it Universidad Complutense de Madrid, 28040 Madrid, Spain.
}

\date{May 25, 1998}

\maketitle

\thispagestyle{empty}

\begin{abstract}
\small
Using Finite-Size Scaling techniques, we numerically show that the
first irrelevant operator of the lattice $\lambda\phi^4$ theory in
three dimensions is (within errors) completely decoupled at
$\lambda=1.0$.  This interesting result also holds in the
Thermodynamical Limit, where the renormalized coupling constant shows
an extraordinary reduction of the scaling-corrections when compared
with the Ising model.  It is argued that Finite-Size Scaling analysis
can be a competitive method  for finding improved actions.
\end{abstract}

\noindent {\it PACS:} 
05.50.+q  
75.40.Mg  
75.40.Cx  
11.15.Ha  

\noindent {\it Keywords:}
Lattice.
Monte Carlo.
Perfect actions.
Critical exponents.
Finite-size scaling.

\section{\protect\label{S_INT}Introduction}

The study of the scale-invariance of a system with many-degrees of
freedom is of relevance both for the study of phase transitions in
Condensed-Matter Physics and for High Energy Physics.  The
Renormalization Group (RG) is the central concept for these
investigations. In both fields it can be turned onto a powerful tool
when combined with Monte Carlo (MC) simulations~\cite{BOOKS}.  In this
kind of investigations one is only interested in universal properties
(in RG language, the leading singularities described by the relevant
operators). However, the action used in the simulation is also coupled
to irrelevant operators.  Their effects can only be neglected in the
limit of infinite lattice size and infinite correlation-length.  This
systematic error is at present the main difficulty to extract
meaningful results from the simulation. For phase-transitions studies,
the quantitative consideration of the scaling-corrections induced by
the irrelevant operators in Finite-Size Scaling (FSS)
studies~\cite{BARBER} was started in Ref.~\cite{BINDER}. At present,
the problem is quite well understood, both
below~\cite{BLOTE,ON,ISINGPERC}, and at the upper critical
dimension~\cite{ISDIL4DJJ}.

An action completely decoupled of the irrelevant operators is called a
perfect action~\cite{WILSON4}.  From the RG point of view, a perfect
action is on the RG-trajectory that leaves the fixed point along the
(supposedly unique) relevant direction (the so-called Renormalized
Trajectory).  It can be easily understood that the effort of finding
such an action may be rewarding. In fact, the quest for perfect
actions has been very active in the last years
(see Ref.~\cite{NIEDHAS} and references therein).  The
perfect action program may be summarized as follows. One first choose
a real-space RG transformation, usually keeping a tunable parameter. A
finite-dimensional coupling space is chosen, and the RG trajectory is
tried to fit in it. The tunable parameter is played with, in order to
minimize the truncation errors due to the finite-dimension of the
coupling-space~\cite{XY}. 

The main difficulty in the perfect action program described above is
related with the arbitrariness introduced by the RG transformation and
by the coupling-space chosen to parameterize it.  Indeed, given a
reasonable coupling-space, it is expected  to
contain a set of points for which scaling-corrections are minimal
(this is a working definition of a ``perfect'' action). It
is clear that the most efficient way of locating it cannot be trying
different RG group transformations until one of them falls just on
it. We will argue that FSS analysis
provides a simple and efficient way to locate this privileged set of
couplings. To put the argument at work, we have investigated the
lattice scalar $\lambda\phi^4$ theory in three dimensions. We will
show that in this simple two dimensional coupling space it is
possible to put the action's coupling to the first irrelevant operator
below the statistical errors.  However, the usual lattice gauge-theories
simulations are not done in the FSS regime (correlation length $\xi$
much larger than the lattice size $L$), although some interesting
calculations have been performed~\cite{LUSHER}.  For this reason, we
have also investigated the scaling properties of our ``optimal''
action in the opposite regime, $L\gg\xi$. We have found again that the
effects of the leading irrelevant operator lies below the statistical
errors. As a byproduct, we are able to reconcile some discrepancies
between MC calculations of the renormalized coupling constant at
zero external momentum~\cite{BAKER,GUPTA}, with a previous
field-theoretic estimate~\cite{BAKER2}.

Very recently a similar MC study has been published~\cite{HASENBUSCH}
for the $S=1$ Ising model. The freedom introduced is quite close to
ours (we let the spin to be a real number, they only allow it to be
$1$,$0$ or $-1$). Thus, it is worthwhile to briefly comment on the
differences between the two works.  To start with, they use as input
the value of a numerically determined universal quantity (basically a
Binder cumulant), in order to numerically tune the action.  On the
contrary, we do not require a previous knowledge of the Binder
cumulant. Next, they use their improved action to produce a strong
error reduction on the estimates of the Ising critical exponents.  We
believe this error reduction to be unjustified (see Ref.~\cite{ISINGPERC},
and section \ref{S_NRFSS}).

\section{\protect\label{FSSLOCATION }Finite Size Scaling and the best
action}

To frame the discussion, let us consider a model with only two
relevant parameters, the ``thermal field'', $t$, and the ``magnetic''
field, $h$, like for instance the Ising model. The free-energy
of a sample of linear size $L$ can be written as~\cite{BARBER,BLOTE}
\begin{equation}
f(t,h,\{u_j\}, L^{-1})=g(t,h,\{u_j\})
+s^{-d} f_{\mathrm {sing}}(s^{y_t} t, s^{y_h} h, 
\{u_j s^{y_j} \}, s/L),\ j\geq 3.
\label{rg_L}
\end{equation}
where $s$ is the scale of an (unspecified) RG transformation, and
$f_{\mathrm {sing}}$ stands for the singular part, $g$ being
analytical. The $y$'s are the eigenvalues of the RG transformation, so
that $y_t,y_h>0$, but $0>y_3>y_4>\ldots$ . A basic assumption of this
approach is that the RG-evolution of the couplings is the same as in
an infinite system.  It is quite common to write $\nu=1/y_1$,
$\eta=2+d-2y_h$ and $\omega=-y_3$. From the free-energy, the
thermodynamically interesting observables follow by taking
derivatives.  Let $\xi(L,t)$ be any reasonable definition of the
correlation length in a finite lattice (see for instance
Ref.~\cite{COOPER}).  One can obtain a general expression for any
observable, $O$ diverging in the thermodynamical limit like
$t^{-x_O}$:
\begin{equation}
O(L,t)=L^{x_O/\nu}\left[ F_O\left(\frac{\xi(L,t)}{L}\right)
+ {\cal O}(L^{-\omega},\xi_\infty^{-\omega})\right]\ ,
\label{FSS}
\end{equation}
where $\xi_\infty$ is the correlation-length in an infinite lattice.
Eq.~(\ref{FSS}) is most interesting deep in the scaling region
($\xi_\infty\gg L$), where the ${\cal O} (\xi_\infty^{-\omega})$
corrections can be safely neglected. It is clear, that in
Eq.~(\ref{FSS}) we have only kept the corrections due to the first
irrelevant field, $u_3$, (in fact, a full series in $L^{-\omega}$ is
to be expected) but similar corrections arise from the others. A
radically different correction is produced by the analytical part of
the free energy, $g$.  When one takes derivatives in Eq.~(\ref{rg_L})
with respect to the ``magnetic'' field, a not diverging term follows
from $g$. This implies that for several important operators like the
susceptibility, the Binder cumulant, or the correlation length, a
correction like $L^{-\gamma/\nu}$ should be added in Eq.~(\ref{FSS}),
if the multiplicative structure is to be kept.

One can exploit Eq.~(\ref{FSS}), by comparing the measures taken in
two lattices
\begin{equation} 
Q_O=\frac{\langle O(sL,t)\rangle}{\langle O(L,t)\rangle},
\end{equation}
at the
value of the ``temperature'', for which the correlation length in units
of the lattice size is the same in both
\begin{equation}
\left.Q_O\right|_{Q_\xi=s}=s^{x_O/\nu}
+A_{Q_O} L^{-\omega}+\ldots\ ,
\label{QUO}
\end{equation}
where $A_{Q_O}$ is a constant. In this way, not only the critical
exponent $x_O/\nu$ can be measured, but also information on the
scaling corrections $A_{Q_O} L^{-\omega}$ can be
extracted~\cite{ON,ISINGPERC,RP2,ISDIL3D,PERC}. Here we encounter an
objective property that a perfect action should have: $A_{Q_O}$ should
be zero for all the observables. Notice that this {\it does not} mean
that there are no scaling corrections. There is no way to avoid the
${\cal O}(L^{-\gamma/\nu})\approx {\cal O}(L^{-2})$ induced by the
analytical part of the free energy (they will be there, even if all
the irrelevant fields in Eq.~(\ref{rg_L}), $u_j$ are set to zero).
But we are still not done, because to efficiently locate the points
for which $A_{Q_O}=0$ something should be known about $x_O/\nu$. The
most simple procedure is to choose for $O$ a quantity having $x_O=0$,
like a Binder cumulant, $B$ (see section~\ref{MODEL} for its
definition).

Therefore, having a simple model, like for instance the Ising model, a
simple strategy for improving its scaling behavior is the
following. The original model has only a tunable coupling $\beta$,
that reaches a critical coupling, $\beta_{\mathrm c}$ (in
Eq.~(\ref{rg_L}) $t$ can be identified with the ``reduced
temperature'' $(\beta-\beta_{\mathrm c})/\beta_{\mathrm c}$). We need
to extend the action with another coupling, $\lambda$, which can be
for instance a next-to-nearest neighbors coupling, a $\lambda \phi^4$
term, etc. The critical point $\beta_{\mathrm c}$ will extend into a
critical line in the ($\beta$,$\lambda$) plane, $\beta_{\mathrm
c}(\lambda)$. If there is something like a RG fixed point in the
($\beta$,$\lambda$) plane, one should have
\begin{equation}
1=\frac{B(\beta,\lambda,sL)}{B(\beta,\lambda,L)},\ {\mathrm {if}}\ 
s=\frac{\xi(\beta,\lambda,sL)}{\xi(\beta,\lambda,L)},
\label{PERFECT}
\end{equation}
which is Eq.~(\ref{QUO}) in the absence of scaling corrections.
We shall show that we can tune $\lambda$ to the condition given
in Eq.~(\ref{PERFECT}), with moderate numerical effort, and that
this yields a quite strong reduction of scaling corrections.

One could object that there is no reason for the Fixed-Point to lie in
such a small coupling-space. The answer is twofold. First, there is
quite clear empirical evidence that this indeed happens\footnote
{From the numerical point of view, a truly perfect action and an
action decoupled from the first irrelevant operator cannot be really
distinguished, since one anyway has tiny but measurable analytical
corrections.}, for instance in the two dimensional Ising
model~\cite{ISDIL2D}, the O(4) Non Linear $\sigma$ Model in three
dimensions~\cite{ON} and in the three dimensional site diluted Ising
model, at $80\%$ spin-concentration~\cite{ISDIL3D}. Second and most
important, we are only demanding that there exists {\it some} RG
transformation whose Fixed-Point is close (in some sense) to the
($\beta$,$\lambda$) plane, but we are not burdening ourselves with the
task of finding it. If it exists, we should be able of finding the
fixed-point coordinates with Eq.~(\ref{PERFECT}). And that can be done
with rather small lattices.

\section{\protect\label{MODEL}The model}

We have considered the scalar $\lambda\phi^4$ theory in a three
dimensional cubic lattice of side $L$, with periodic boundary
conditions. The action is given by
\begin{equation}
S=\beta\,\sum_{<i,j>} \phi_i \phi_j\ +\ 
\sum_i \phi_i^2\ +\ \lambda \sum_i \left(\phi_i^2\,-\,1\right)^2 \, ,
\label{ACCION}
\end{equation}
where the $\phi$'s are real variables.  The $\lambda\to\infty$ limit
of the model (\ref{ACCION}) is the Ising model for magnetism. At fixed
$\lambda$, for low values of $\beta$ the system is in a
$Z_2$-symmetric (paramagnetic) phase. There is a critical line,
$\beta_{\mathrm c}(\lambda)$, separating the paramagnetic phase from
the ferromagnetic phase. The full critical line is second order, and
it belongs to the Ising Universality Class, excepting the $\lambda=0$
end-point which is the Gaussian model.  The MC simulation of
(\ref{ACCION}) is greatly eased by the
cluster-methods~\cite{CLUSTMET}.

The observables to be measured are easily defined in terms of the
Fourier transform of the spin-field
\begin{equation}
\widehat\phi(\bbox{k})=\frac{1}{V}\,\sum_{\sbbox{x}}
\phi(\bbox{x})\e^{{\mathrm i}\sbbox{k}\cdot\sbbox{x}}\, ,
\end{equation}
where $V=L^3$ is the lattice volume. We have the susceptibilities at
zero ($\chi$) and minimal momentum ($F$):
\begin{equation}
\chi=V\,\left\langle |\widehat\phi(0)|^2\right\rangle\ ,\ 
F=\frac{V}{3}\,\sum_{\Vert\sbbox{k}\Vert =\frac{2\pi}{L}}
\left\langle|\widehat\phi(\bbox{k})|^2\right\rangle\, ,
\label{SUSCEP}
\end{equation}
and the finite-lattice correlation-length~\cite{COOPER} and
Binder cumulant.
\begin{equation}
\xi=\sqrt{\frac{\chi/F\, -\, 1}
{4 \sin^2 (\pi/L)}}\ ,\ B=\frac{3}{2}-
\frac{\left\langle |\widehat\phi(0)|^4\right\rangle}
{2\left\langle |\widehat\phi(0)|^2\right\rangle^2}\, .
\label{XIBINDER}
\end{equation}
The renormalized coupling constant at zero external momentum can be readily
calculated~\cite{COOPER}
\begin{equation}
g_{\mathrm{R}}=2\, B (L/\xi)^D\ ,\ D=3.
\end{equation}
We also measure the nearest-neighbor energy
\begin{equation}
E=\left\langle\sum_{<i,j>} \phi_i \phi_j\right\rangle,
\end{equation}
in order to calculate $\beta$-derivatives and extrapolations~\cite{REWEIGHT}.
Let us summarize the critical behavior of the
different observables:
\begin{equation}
{\begin{array}{clcccl}
x_{\xi}&=&\nu\, ,&
x_{\chi}&=&\gamma=\nu (2-\eta)\, , \\
x_{g_{\mathrm R}}&=&0\, ,&
x_{\partial_\beta \xi}&=& 1 + \nu\, ,
\end{array}}
\end{equation}
the critical exponents being~\cite{ISINGPERC}
\begin{equation}
\nu=0.6294(10)\ ,\ \eta=0.0374(12)\ ,\ \omega=0.87(9).
\label{ISINGEXP}
\end{equation}
These MC results are in excellent agreement with the last series 
estimates~\cite{ZINNJUSTIN}.

A final remark is that ~\cite{BAKER}
\begin{equation}
g_{\mathrm R}^\infty\equiv\lim_{\beta\to\beta_{\mathrm{c}}^-} 
\lim_{L\to\infty}\, g_{\mathrm R}(L,\beta)\neq
\lim_{L\to\infty}\lim_{\beta\to\beta_{\mathrm{c}}}\,
g_{\mathrm R}(L,\beta)\equiv \tilde g_{\mathrm R} \, , 
\label{GRS}
\end{equation}
as Eq.~(\ref{FSS}) shows. 

\section{\protect\label{OPTIMAL} Optimizing $\lambda$}

For a first estimate of the Fixed-Point location, we have simulated
lattices $L=4,8$ and $16$ in a mesh of $\lambda$ values
($\lambda=0.25,0.5,1.0,1.5$ and $2$).  Results in the Ising limit were
also available from previous work~\cite{ISINGPERC}.  We have measured
$\left.Q_B\right|_{Q_\xi=2}$ for lattice-pairs $(4,8)$ and ($8$,$16$),
and check for the condition in Eq.~(\ref{PERFECT}) . Given the strong
statistical correlation between $Q_\xi$ and $Q_B$, this quotient can
be very precisely calculated. One can obtain a $0.2\%$ accuracy in the
($8$,$16$) pair, with nine Pentium Pro hours. The original idea was to
use the mesh as starting point for a bisection search of the optimal
$\lambda$, but the $\lambda=1$ data fulfilled the condition in
Eq.~(\ref{PERFECT}) to the achieved accuracy. Moreover, for the
lattice pair ($8$,$16$), at $\lambda=1.5$, $Q_B$ was at three standard
deviations below 1, and at $\lambda=0.5$ was five standard deviations
above.  From this, one can estimate the optimal $\lambda$ to be
$1.0(1)$.  The total CPU time was less than two hours of the 32
Pentium Pro machine RTNN.

A more precise location of the optimal $\lambda$ would require not
only much more statistics, but also to simulate larger lattices to
avoid higher-order scaling corrections (if needed, one could resort to a
reweighting method in $\lambda$). Nevertheless our main scope,
has been to check if a zero value of $A_{Q_B}$ (see Eq.~(\ref{QUO}))
does imply the vanishing of the ${\cal O}(L^{-\omega})$ corrections
for the other quantities of interest.  In other words, if the first
irrelevant operator is decoupled.  For this, we have simulated at
$\lambda=1$, and also at $\lambda=0.5$ and $\lambda=0.25$ for
comparison. The total CPU time employed has been about three RTNN
months. We remark that such a time-consuming investigation is not
an essential part of an action improvement program, but rather
a consistency check. 

In practice, one is mainly interested in the thermodynamical limit,
where the measures are quite less accurate than in the FSS region.
For instance one could be interested in measures of exponential tails
of propagators, or in measures of renormalized coupling constants. In
the model considered in this work maybe the more interesting universal
quantity is $g_{\mathrm R}^\infty$.  Even with the high accuracy
allowed by the improved cluster estimators~\cite{WOLFF2}, we will show
that a $10\%$ error in the optimal $\lambda$ is enough to cancel the
leading scaling-corrections. In fact the $\xi=2.5$ data are in the
continuum limit to our accuracy ($2\%$). In this regime, we have used
6 RTNN days.

\section{\protect\label{S_NRFSS}Numerical Results when $\xi_\infty\gg L$}

\begin{table}[b!]
\begin{center}
\begin{tabular*}{\linewidth}{@{\extracolsep{\fill}}rlllll}\hline\hline
\multicolumn{1}{c}{$L$} 
& \multicolumn{1}{c}{Ising} 
& \multicolumn{1}{c}{$\lambda=1$} 
& \multicolumn{1}{c}{$\lambda=0.5$} 
& \multicolumn{1}{c}{$\lambda=0.25$} \\ \hline
8    & 0.98146(20) &1.00094(25)&1.01372(30)&1.02949(40)\\
12   & 0.98703(23) &1.00075(24)&1.01088(32)&1.02251(35)\\
16   & 0.99038(23) &1.00062(26)&1.00817(29)&1.01840(38)\\
24   & 0.99324(21) \\
32   & 0.99476(28) &0.9997(5)&1.0045(5)&1.0095(7)\\
48   & 0.99612(28) \\
64   & 0.99702(34) \\\hline
\end{tabular*}
\caption{Quotient of the Binder cumulant measured in lattice
pairs ($L$,$2L$) where the correlation-lengths in units of the
lattice size are the same. Scaling corrections appear as
a value different from one.}
\label{LAMBDA}
\end{center}
\end{table}

\begin{figure}[t!]
\begin{center}
\leavevmode
\epsfig{file=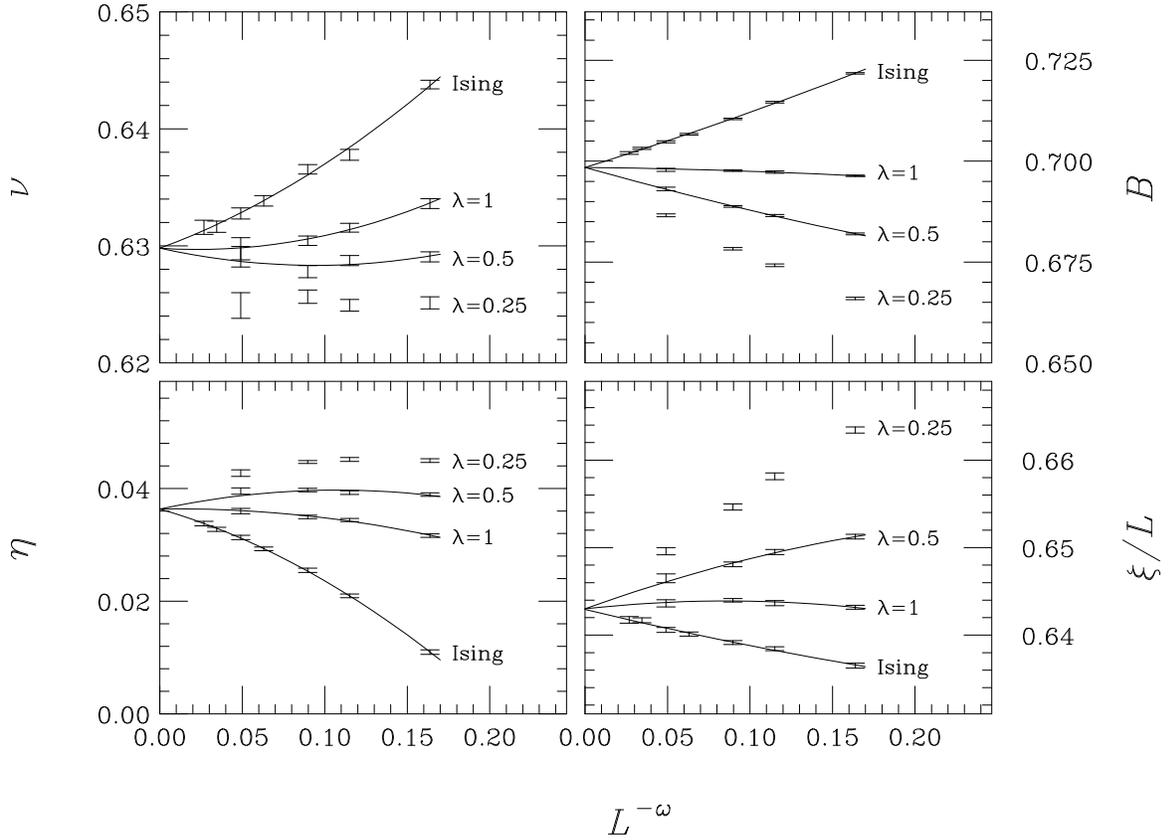,width=0.7\linewidth,angle=90}
\end{center}
\caption{Universal quantities as a function of $L^{-\omega}$
($\omega=0.87$), for
the four models considered. The lines correspond to quadratic fits, 
constrained to yield the same infinite-volume extrapolation.
}
\label{EXPO1}
\end{figure}

\begin{figure}[t!]
\begin{center}
\leavevmode
\epsfig{file=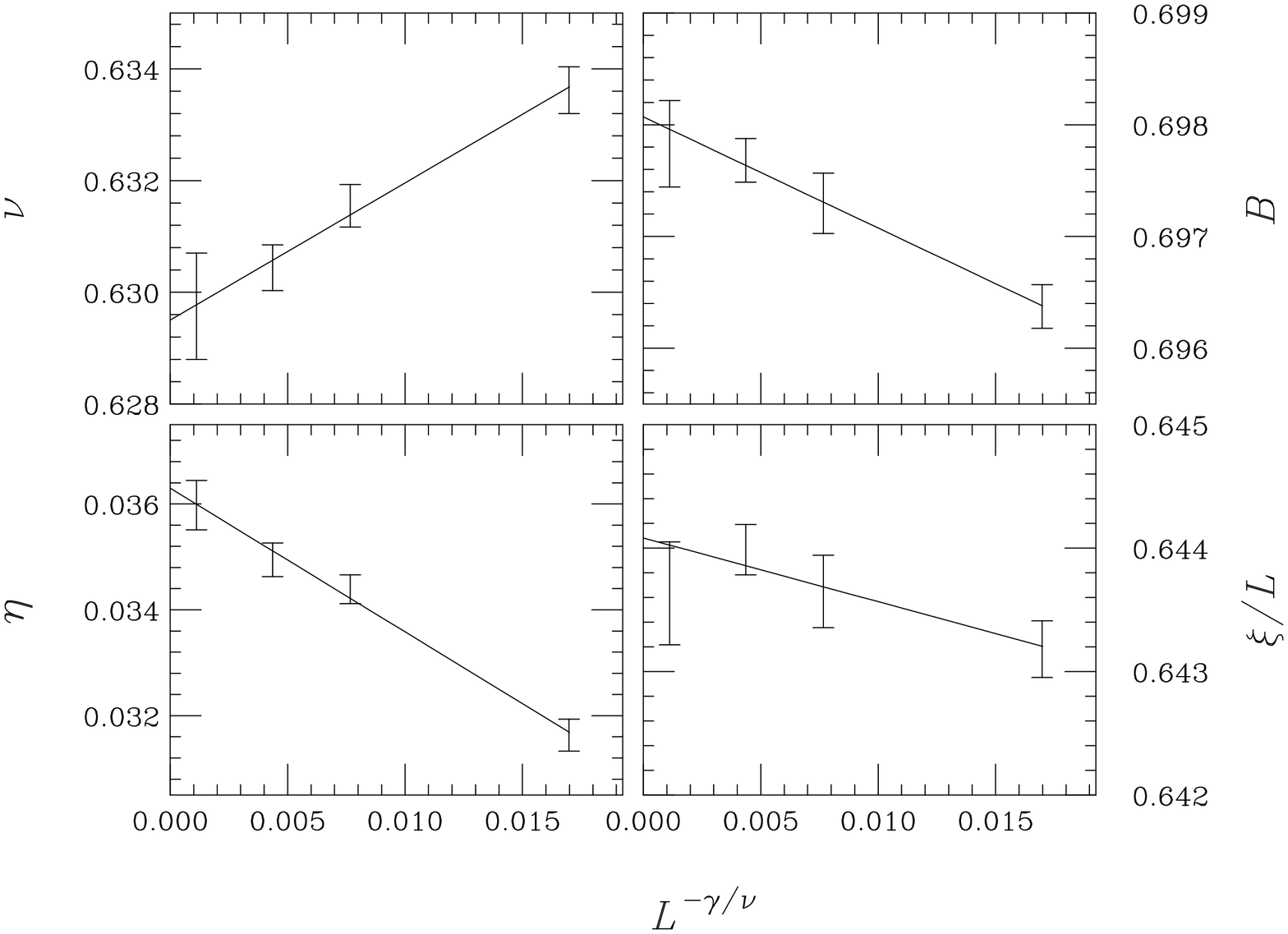,width=0.7\linewidth,angle=90}
\end{center}
\caption{Universal quantities as a function of $L^{-\gamma/\nu}$, for
the $\lambda=1$ data. The linear behavior shows that there is no
evidence for scaling-corrections others than the analytical ones.}
\label{EXPO2}
\end{figure}

We have simulated in lattices $L=8,12,16,24,32$ and $64$ for
$\lambda=1.0,0.5$ and $0.25$. The Elementary Monte Carlo Step (EMCS)
was composed of 10 single-cluster flips, followed by a full-lattice
Metropolis sweep.  We have taken $10^7$ measures, separated by four
EMCS, excepting $L=64$ where only $2.5\times10^6$ measures were
taken. Our values for $\left.Q_B\right|_{Q_\xi=2}$ are quoted in table
\ref{LAMBDA}.  From Eq.~(\ref{QUO}) we can obtain an ($L$ dependent)
estimate of $\nu$ and $\eta$, and also of
$\left.B(L)\right|_{Q_\xi=2}$ and
$\left.\frac{\xi}{L}\right|_{Q_\xi=2}$.  The results are displayed in
Fig.~\ref{EXPO1}, where a quadratic fit in $L^{-\omega}$ is
presented. The results for the linear term in $L^{-\omega}$ of the
$\lambda=1$ data are (fitting independently from the other $\lambda$
values)
\begin{equation}{\begin{array}{cccccc}
A_{Q_{\partial_\beta\xi}} &=& 0.004(44)\ ,&
\ A_{Q_{\chi}}&= &0.016(25)\, ,\\
A_{\xi/L}&=&0.022(23) \ ,&\ A_{B}&=&0.010(21)\, .
\end{array}}
\label{OK}
\end{equation}
Thus, it seems quite clear that for the four quantities shown in
Fig.~\ref{EXPO1}, the coefficient of the ${\cal O}(L^{-\omega})$
corrections changes sign close to $\lambda=1.0$ and that the
$\lambda=1.0$ data are compatible with a ``perfect'' behavior. That
would mean that the scaling-corrections may arise from other
irrelevant operator, and from the analytical corrections ${\cal
O}(L^{-\gamma/\nu})$. Plotting our data against the analytical
corrections, a nice linear behavior is found (see
Fig.~\ref{EXPO2}). Therefore, to our precision $\lambda=1$ seems
perfectly decoupled from the first irrelevant operator. Operators with
$\omega'>\omega$ cannot be resolved from the analytical corrections.
At this point, one could be tempted of using the fits of
Fig.~\ref{EXPO2} to reduce the error in the infinite volume
extrapolation for the critical exponents, as it is done for instance
in Ref.~\cite{HASENBUSCH}.  Followed blindly, this criterion reduces
the errors in a factor four, or larger when compared to the fit in
Fig.~\ref{EXPO1}.  But we believe this reduction to be unjustified,
because we have tuned $\lambda$ to Eq.~(\ref{PERFECT}) {\it
numerically}, so the condition $A_{Q_O}=0$ can only be expected to
hold {\it within errors}, as shown in Eq.~(\ref{OK}). Therefore, the
more conservative estimate given in Eq.~(\ref{ISINGEXP}) is to be
preferred. If a procedure were designed (numerical or analytical) to
largely reduce the errors in $A_{Q_O}$, one could expect an
improvement of an order of magnitude in critical exponents measures.
Nevertheless, we will see below that a truly significant gain will be
obtained in the $L\gg\xi$ regime.

A worrying feature of this analysis, is that the scaling corrections
for $\eta$ and $\nu$ when $\lambda=0.25$, seem no longer under
control. One would say that we need to simulate in significantly
larger lattices, to reach the asymptotic regime. We think this to be
an effect of the competition with the Gaussian fixed-point at
$\lambda=0$.  A similar behavior was found in the site diluted Ising
model at very small dilution~\cite{ISDIL3D}.

\section{\protect\label{S_NRFLT}Numerical Results when $\xi\ll L$}

\begin{figure}[t!]
\begin{center}
\leavevmode
\epsfig{file=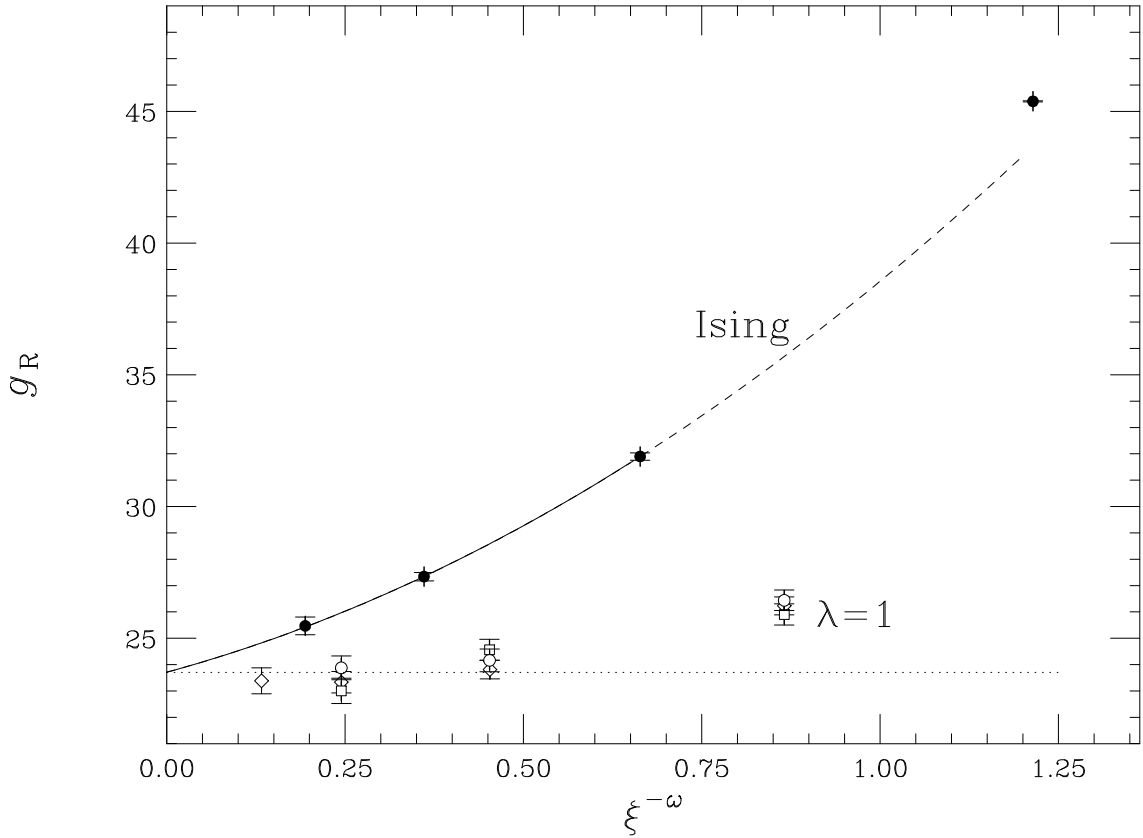,width=0.7\linewidth,angle=90}
\end{center}
\caption{Renormalized coupling constant for $\beta<\beta_\mathrm{c}$.
The Ising data are taken from Ref.~\cite{BAKER} and correspond to
$L/\xi=10$. The $\lambda=1$ data are obtained at $L/\xi=6$ (diamonds),
$L/\xi=10$ (squares) and $L/\xi=12$ (circles). The horizontal dotted
line correspond to the theoretical value.
}
\label{GR}
\end{figure}

As an objective test of scaling, we have chosen the renormalized
coupling constant. This has been a remarkably difficult quantity
to measure until now. Field Theoretic calculations~\cite{BAKER2}
yielded
\begin{equation}
g_{\mathrm R}^\infty=23.73(2),
\label{GRFT}
\end{equation}
which was difficult to reconcile with MC calculations up to
correlation-length $\xi=6$~\cite{BAKER}. Moreover,
calculations~\cite{GUPTA} of $\tilde g_{\mathrm R}$ (see
Eqs.~(\ref{GRS})) show a worrying decreasing tendency, even suggesting
a hyper-scaling violation. Our data in the right-hand side of
Fig.~\ref{EXPO1} quite convincingly show that both the Binder Cumulant
and $\xi/L$ are non zero and universal at criticality.  Thus, $\tilde
g_{\mathrm R}$ is certainly not zero, as expected. In fact, using the
data and procedure of~\cite{ISINGPERC}, we obtain
\begin{equation}
\tilde g_{\mathrm R}=5.25(3)
\end{equation}
However, the more interesting
quantity to calculate is $g^\infty_{\mathrm R}$ in the reversed limits
ordering. The theoretical expectation~\cite{WEGNER} is that scaling
corrections will appear as a power series in $\xi_\infty^{-\omega}$.

We have simulated with $\lambda=1.0$, for correlation-lengths
$\xi=1.25,2.5,5$ and $\xi=10$, that correspond to
$\beta=0.316,0.3588,0.3729$ and $0.3776$, respectively. The EMCS
consists of a Swendsen-Wang update of the spin signs, followed by a
Metropolis sweep for changing the modulus. Measures were taken every
EMCS, using improved estimators~\cite{WOLFF2} (the CPU gain using this
estimators is not smaller than a factor of $100$, if the need for
$\beta$ reweighting is not as critical as in the FSS region).  We have
carried out $1.5\times 10^6$ EMCS for each lattice size and $\beta$
value. In all cases, but $\xi=10$, we have simulated lattices with
$L/\xi\approx 6,10$ and $12$ (we have used lattices
$L=8,12,16,24,32,48$ or $64$, depending on $\xi$). The finite-size
corrections are expected to drop exponentially with $L$ if periodic
boundary conditions are used. At our accuracy level ($2\%$ in the
renormalized coupling constant), no significant finite-size
corrections are found for $L/\xi\geq 6$ (see Fig.~\ref{GR}). Thus, for
$\xi=10$, we have only simulated a $L=64$ lattice.

Our data are plotted together with those for the Ising model of
Ref.~\cite{BAKER}, which have been taken at $L/\xi=10$, thus being
quite asymptotic. The horizontal dashed line is the continuum-limit
value of Eq.~(\ref{GRFT}).  For the Ising model we draw a quadratic
fit in $\xi^{-\omega}$ (with $\omega=0.87$), constrained to pass
through the value in Eq.~(\ref{GRFT}) (horizontal dotted line),
including only data with $\xi>1$.  The Ising data at $\xi=0.8$, is
somehow far from the quadratic fit, but for $\xi<1$ it is not really
surprising that one needs several terms of the $\xi^{-\omega}$
series. In the fit, a significantly non zero coefficient for
$\xi^{-\omega}$ is found, as one would expect, and also an important
quadratic contribution.

Regarding the $\lambda=1$ data, at $\xi=1.25$, the scaling-corrections
are at the $10\%$ level (to be compared with a $50\%$ in the Ising
model).  From the data we observe a cancellation (within errors) of the
linear term and a significant reduction of the quadratic corrections.
This could be interpreted as a full cancellation of the $\xi^{-\omega}$
series (full decoupling of the first irrelevant operator).  As for
$\xi\geq 2.5$ our results are compatible with the Field-Theoretical
value, we cannot distinguish if the deviations at $\xi=1.25$ are due
to other irrelevant operators or to the analytical corrections (which
would be ${\cal O}(\xi^{-\gamma/\nu})$).

As a final comment, the $\lambda=1.0$ two-point correlator, $\langle
\phi(x)\phi(y)\rangle$, does not show a significant improvement on its
rotational-invariance when compared with the Ising correlator. That is
not really surprising, because we have included only a first-neighbors
coupling. So, at correlation-length $\xi\approx 1$ we only have a
statistical system with highly anisotropic couplings, which is
naturally reflected in the high-momentum behavior of the propagator.

\section{\protect\label{S_CON}Conclusions}

We have shown that by eliminating the scaling corrections of the
scaling-function of the Binder cumulant in the FSS regime, we obtain a
radical improvement of the scaling behavior of other critical
quantities, both in the FSS regime and in the thermodynamical limit.
This has been done by considering objective properties of the system
under study, not depending on an {\em ad hoc} RG transformation.  This
result poses at least three questions. How critical is the choice of
the tunable parameter, $\lambda$? (that is, could we have obtained
such a good result by considering a next-to-nearest neighbors
coupling, for instance?) How can this optimization strategy be
implemented in an asymptotically-free theory? (this could be
investigated in the two dimensional, $O(3)$ Non-Linear $\sigma$ model,
and will be considered in a near future) And last, but not least, will
this strategy work in an asymptotically-free lattice-gauge theory?

Regarding the reduction of the errors, the optimal value of $\lambda$
is only approximately known, thus the ${\cal O}(L^{-\omega})$
corrections cannot be completely disregarded.  As a consequence, not
true gain in the critical exponents determination is achieved.  The
situation is quite better in the thermodynamical regime ($L\gg\xi$),
where continuum-results (at the $2\%$ level) for the renormalized
coupling constant can be obtained at correlation-length $\xi=2.5$.  Of
course one could object that a judicious use of the FSS
method~\cite{FSSAF} allows to obtain thermodynamical data at very
large correlation-length. But also in this case there is a real danger
that scaling-corrections will spoil the extrapolation, so a drastic
corrections reduction will be of benefit.

\section*{Acknowledgments}

We thank Giorgio Parisi for an encouraging discussion on the
feasibility of this work. We are also grateful to Juan Jes\'us 
Ruiz-Lorenzo for many interesting discussions and comments.

This work has been partially supported by CICyT (contracts AEN97-1708,
AEN97-1693). The simulations have been carried out in the RTNN
machines at Zaragoza and Complutense de Madrid Universities.

\end{document}